\documentclass[10 pt, conference]{IEEEtran}
\usepackage{epsfig,graphicx,color,amsmath,amsfonts,bm,balance,tikz,array,multirow,fancyhdr}
\usepackage{url}
\usepackage{amsmath}
\usepackage{float}

\title{\LARGE \bf
Centimeter-Level Indoor Localization using Channel State Information with Recurrent Neural Networks
}

\author{Jianyuan Yu and R. Michael Buehrer \\
Bradley Department of Electrical and Computer Engineering, Virginia Tech, Blacksburg, VA 24061 \\
\{jianyuan, buehrer\}@vt.edu
}

\begin{document}

\maketitle
\thispagestyle{empty}
\pagestyle{empty}

\begin{abstract}

Modern techniques in the Internet of Things or autonomous driving require more accuracy positioning ever. Classic location techniques mainly adapt to outdoor scenarios, while they do not meet the requirement of indoor cases with multiple paths. Meanwhile as a feature robust to noise and time variations, Channel State Information (CSI) has shown its advantages over Received Signal Strength Indicator (RSSI) at more accurate positioning. To this end, this paper proposes the neural network's method to estimate the centimeter-level indoor positioning with real CSI data collected from linear antennas. It utilizes an amplitude of channel response or a correlation matrix as the input,  which can highly reduce the data size and suppress the noise. Also, it makes use of the consistency in the user motion trajectory via Recurrent Neural Network (RNN) and signal-noise ratio (SNR) information, which can further improve the estimation accuracy, especially in small datasize learning.  These contributions all benefit the efficiency of the neural network, based on the results with other classic supervised learning methods.

\end{abstract}


\section{INTRODUCTION}

\begin{figure}[h!]
  \centering
  \includegraphics[width=0.4\textwidth]{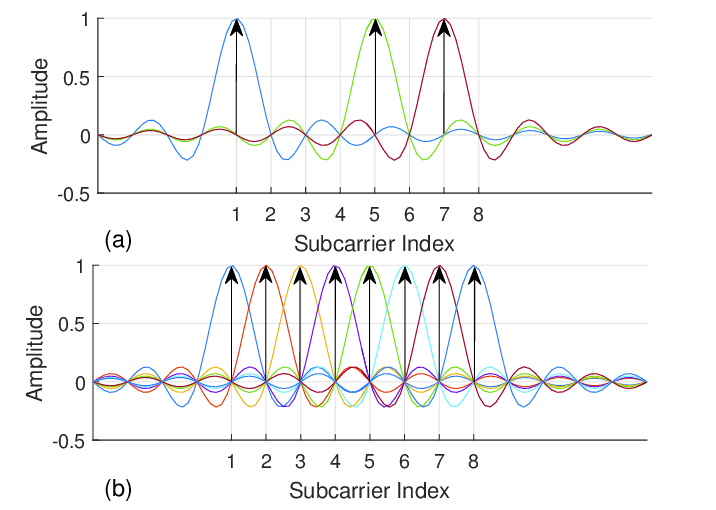}
  \caption{OFDM Subcarriers samples to represent channel state response}
  \label{fig:csi}
\end{figure}

Indoor positioning is a key enabler for a wide range of applications, including navigation, smart factories and cities, surveillance, security, IoT, and sensor networks. Additionally, indoor positioning can be leveraged for improved beamforming and channel estimation in wireless communications \cite{arnold2019novel}. Two major issues in positions are features extracting and fingerprint matching method. For the features, RSSI has been widely used to reach sub-room or sub-meter level accuracy ~\cite{hoang2019recurrent} ~\cite{hsieh2018towards}, but it suffers from the time-varying information and unstable to noise. To this end, Channel State Information (CSI) becomes popular recently due to its robust to noise and time variance. As shown in Fig,~\ref{fig:csi}, the peak of each subcarrier, in different colored lines, in an OFDM system is taken as the CSI, it is considered as the downsampling of frequency domain of channel response ~\cite{dougan2018ofdm}. Earlier work of ~\cite{wu2012csi}  has shown the CSI-based fingerprinting method can offer a more accurate estimation than the sub-meter RSSI method. Besides localization, CSI is also widely used for user activity recognition and classification ~\cite{chowdhury2017wihacs}, and more recent work ~\cite{gringoli2019free} has release CSI collecting tool for users to play with. \\
\indent For the fingerprint matching,  earlier work includes k-nearest neighbors ~\cite{battiti2002neural} and support-vector machine ~\cite{chowdhury2017wihacs}, but they both suffer from overfitting. Random forest ~\cite{jedari2015wi} is an alternative solution, but it may miss important information of input ~\cite{nitze2012comparison} and not accurate enough compared to the neural network. Therefore,  more neural network technics are discussed, the paper ~\cite{laoudias2009indoor} applies Radius Basis Function as the hidden layer to deal with the RSSI dataset in the Wifi environment, and ~\cite{altini2010bluetooth}  implement it in Bluetooth environment, with the enhancement of navigation tree.  What's more, regarding the data is generally collected by the slow-moving user, such characteristics could also be utilized to improve the estimation. ~\cite{abdelbar2017indoor} estimate the location by previous user trajectory.  Recurrent Neural Network is applied in work ~\cite{hoang2019recurrent} ~\cite{hsieh2018towards} using WiFi-RSSI data to better estimate positions. However, among all these RSSI-based methods, the best accuracy is limited due to the time-variant property of RSSI. In particular,  the first prize winner has recently released his solutions dealing with the same open dataset in our work, the author ~\cite{sobehy2019ndr} uses polynomial regression to smooth the data, and further refines the estimation by data augment and ensemble neural network ~\cite{sobehy2019csi}. However, the work does not make full use of the user trajectory and SNR information, either compare it to the decision-tree-based method. \\
\indent In this paper, we propose the  Recurrent Neural Network to estimate the position with CSI data. We start by discussing the feature extraction with a brief validation. Then we introduce several estimation methods and their advantages. Finally, we evaluate the performance of different methods on a different dataset, and explain how the RNN can utilize the user trajectory and SNR information and the impact on estimation accuracy when using either input or method. This work can be further applied in multiple objects positioning or tracking in warehouses, shopping malls or smart factories, where it worth the effort to generate training dataset and labels, and the trained model is heavily used in prediction later.

\section{Problem Formation}

\begin{figure}[h!]
  \centering
  \includegraphics[width=0.4\textwidth]{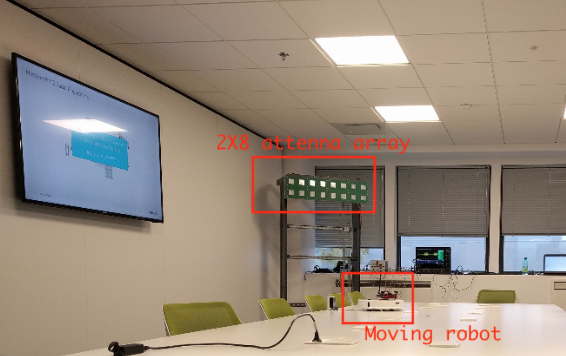}
  \label{fig:setup}
  \caption{Scenario setup， a moving robot on desk communicate with the $2\times8$ uniform linear array}
\end{figure}

The problem is a positioning algorithm competition at the 2019 IEEE Communication Theory Workshop. The object of the competition is to design and train an algorithm that can determine the position of a user, based on estimated \textit{channel frequency responses} between the user and an antenna array. As shown in Fig.~\ref{fig:setup} of work \cite{arnold2019novel}, channel responses were measured between a moving transmitter and an $8\times2$ antenna array. As a transmitter, an SDR-equipped vacuum-cleaner robot was used, and it drove in a random path on a $4\times2$ meter table and transmitted uplink OFDM pilots with a bandwidth of 20 MHz and 1024 subcarriers at a carrier frequency of 1.25GHz. 10 of the subcarriers were used as guard bands. The robot move in a planned path and some other sensors can provide ground truth location. Channel vectors from a dataset created with the channel sounder in described ~\cite{arnold2019novel}  will be used. The dataset comprises channel responses, associated position ground truth information, and  SNR information. The channel response is in the dimension $ Ndat \times$ 16 $\times 924$, where $Ndat$ is the number of measurement points 17486, 16 is the number of antennas and 924 is the number of used subcarriers. The position is $ Ndat \times 3$ the coordinates $[x, y, z]$ yet $z$ is constant.  While the SNR size is  $N_{dat}\times 16$. 

\section{Features Extraction}


\subsection{Subcarriers Samples and Smoothing}

\begin{figure}[h!]
  \centering
  \includegraphics[width=0.4\textwidth]{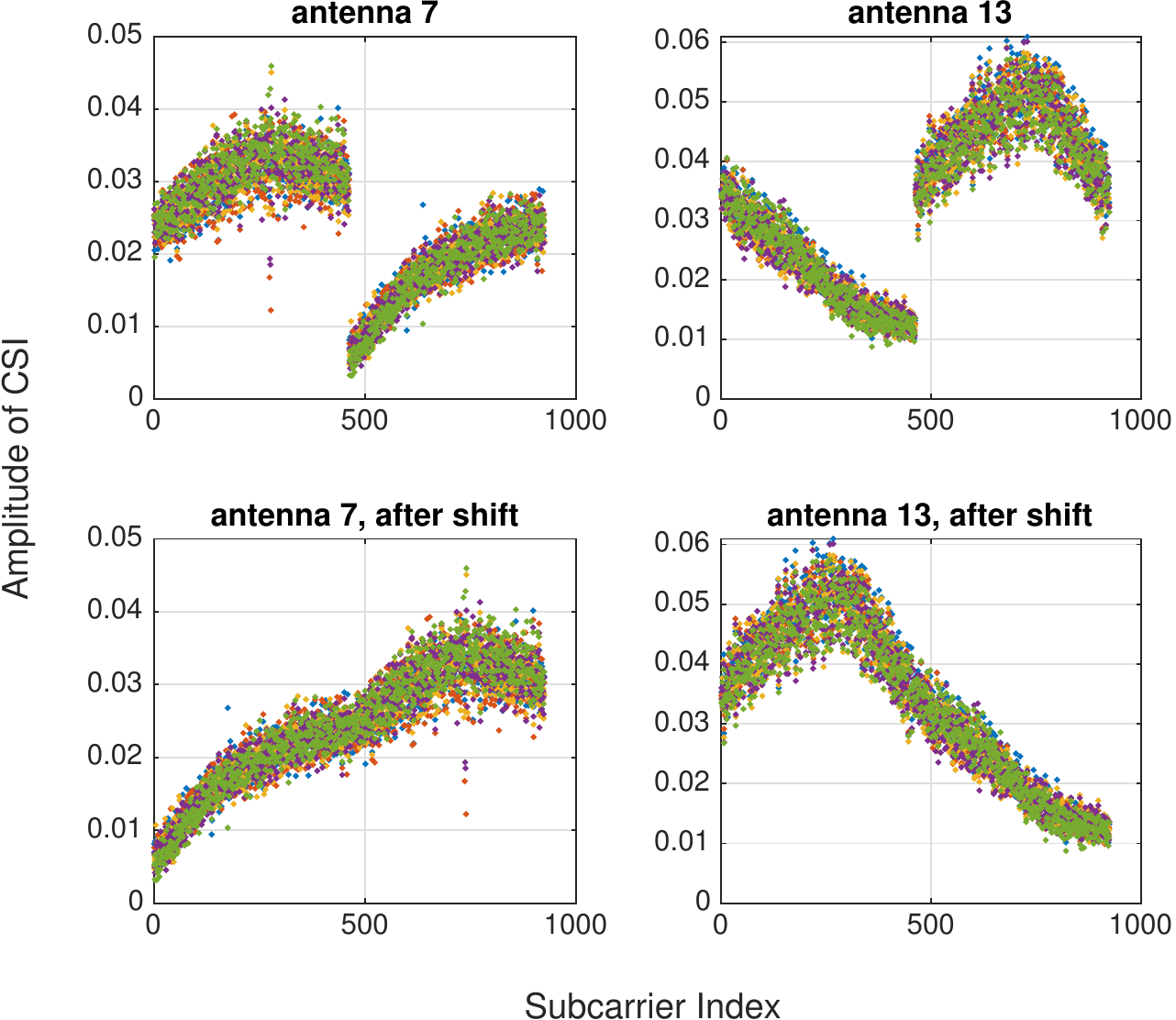}
  \caption{Subcarrier amplitude at 7th and 10th antenna, before and after shifting the data}
  \label{fig:amp_shift}
\end{figure}

The first step is to check the integrity of the data, as plotting the subcarrier amplitude, we fond there is a significant disconnection in the subcarrier data. Since only 10 of 1024 are used as gap subcarriers, the reason is likely the data was shifted when stored. As a solution, we switch the end half as the beginning half, and found that they connect well. Fig. ~\ref{fig:amp_shift} shows the shift operation, where the left column is the plot before the shift, and the right column is after shift.

\begin{figure}[h!]
  \centering
  \includegraphics[width=0.4\textwidth]{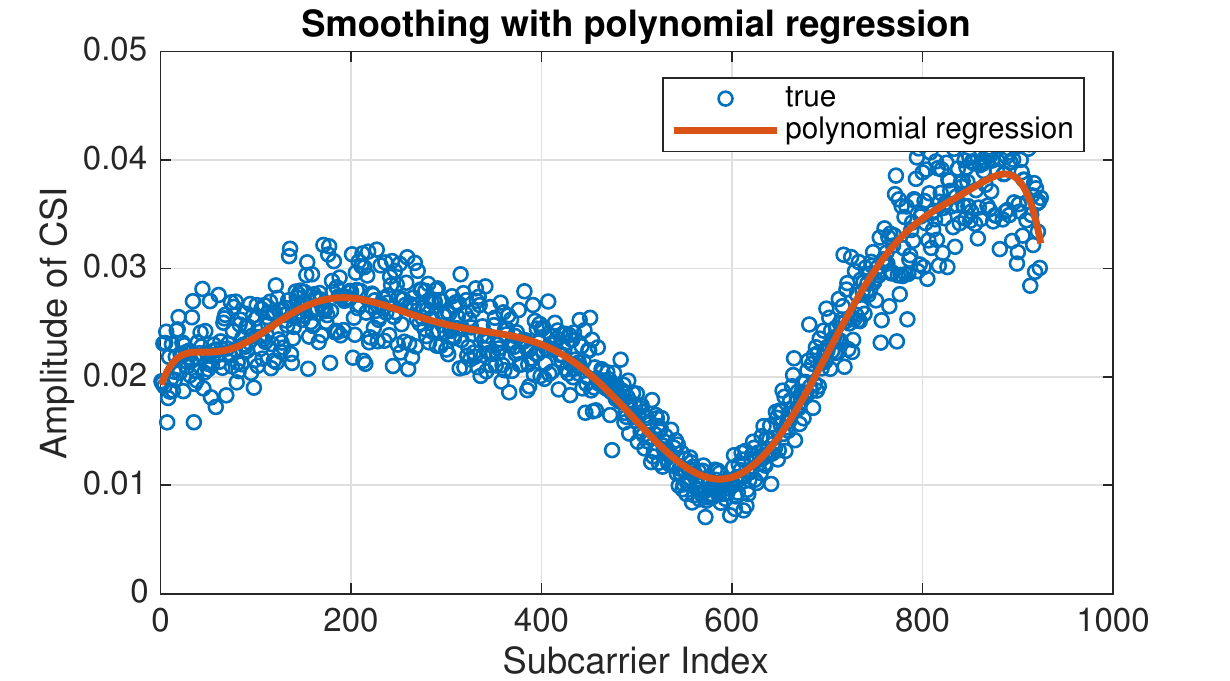}
  \caption{Illustration of  polynomial regression on channel response}
  \label{fig:amp_poly}
\end{figure}
\begin{figure}[h!]
  \centering
  \includegraphics[width=0.4\textwidth]{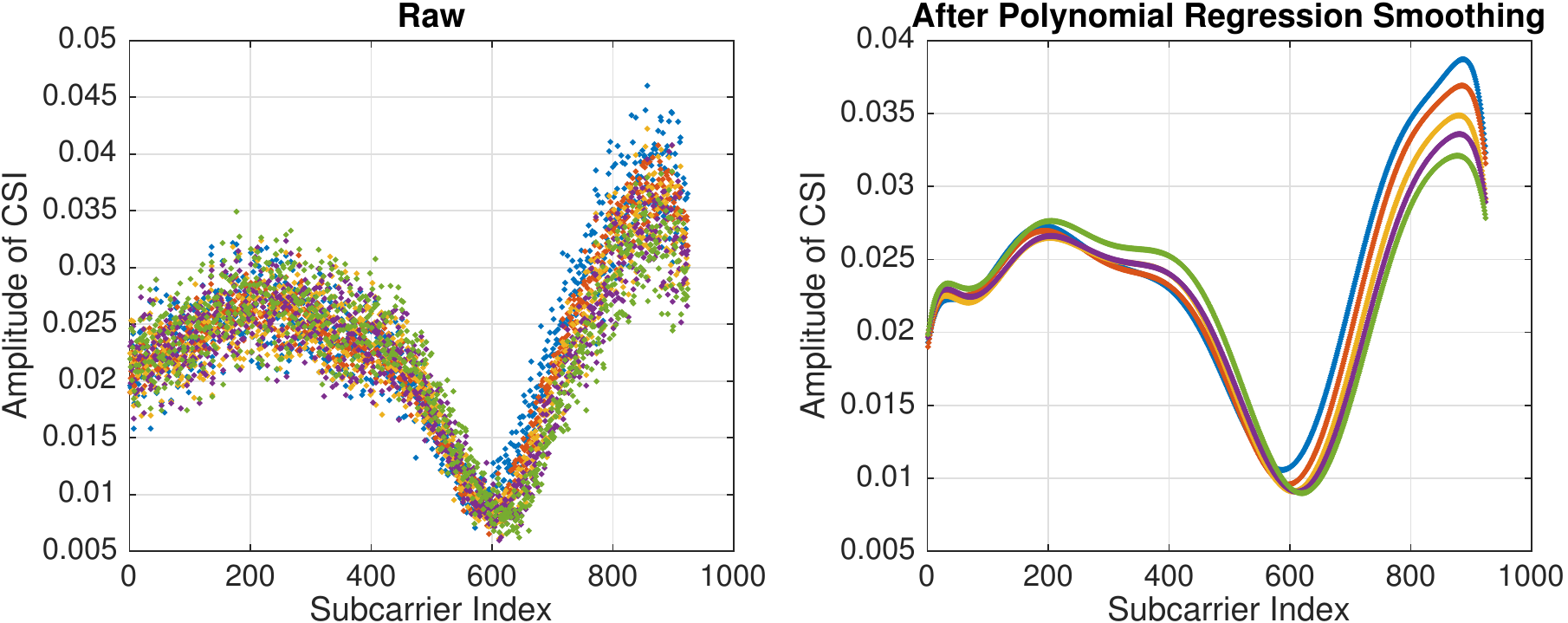}
  \caption{Channel state response before and after polynomial smoothing}
  \label{fig:amp_smooth}
\end{figure}

We reproduce the polynomial regression method ~\cite{sobehy2019ndr} to denoise or smooth the curve of the channel response. The principle of \textit{multiple variable polynomial regression} is to find coefficients $a_i$ that makes $\hat{y} = \sum_{i=0}^{K} a_i x^i$ approximate to $y$ over all time steps, where $y$ is the CSI,  $x$ is time step from 1 to 924, $K$ is the \textit{degree}. It is based on the theory that the curve can be approximately represented by the first few Taylor series. Fig.~\ref{fig:amp_poly} shows an example of the polynomial regression with a degree of 10, and the approximate curve is taken as the channel response after downsampling, it is named \textit{smoothing} since the approximate curse filter out a lot outliner while still maintain the shape of raw data, and keep nearly same shape for nearby position of a fixed antenna. Fig.~\ref{fig:amp_smooth} shows the impact of smoothing,  where these 5 lines are close to each other, at neighboring positions of a fixed antenna.

\subsection{Covariance Matrix (CM)}

\begin{figure}[h!]
  \centering
  \includegraphics[width=0.4\textwidth]{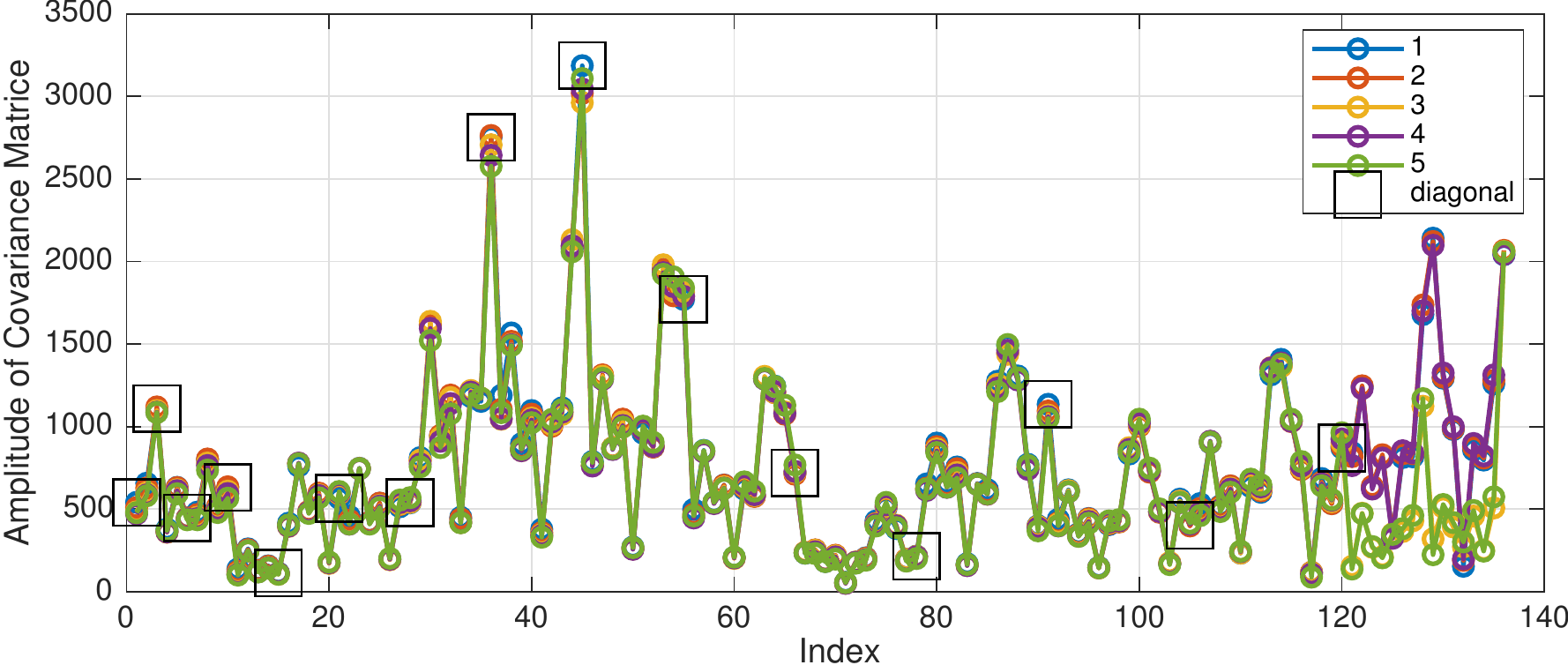}
  \caption{Covariance matrix at same position, the point in square denote the diagonal line of the matrix}
  \label{fig:cm}
\end{figure}

Another feature widely used in direction of arrival estimation is covariance matrix $\textbf{Z}_{M\times M}$, which is the self product of received signal arrays $\textbf{Y}_{M\times T}$, $ \textbf{Z} = \textbf{Y}\times \textbf{Y}^H$, where $M =16$ is the number of antennas, and $T$ is the time steps. Based on $\textbf{Y} = H(\textbf{AS}) + n = \textbf{A}H(\textbf{S}) + n$, where $\textbf{A}_{M \times K}$ is the manifolder only decided by the direction of angles with $K$ as number of target , $H(.)$ is the channel model, $\textbf{S}_{K\times T}$ is digital signals and $n$ is white noise. Therefore we have 
\begin{equation}
\begin{split}
\textbf{Z} &= (\textbf{A}H(\textbf{S}) + n)(\textbf{A}H(\textbf{S}) + n)^H \\
   & = \textbf{A} (H(\textbf{S}) H(\textbf{S})^H)\textbf{A}^H + \delta^2 \textbf{I}. \\
\end{split}
\end{equation}
If the channel is ideal, then we have $H(\textbf{S}) = \textbf{S}$, and $\textbf{Z} = \textbf{A} (\textbf{SS}^H)\textbf{A}^H + \delta^2 \textbf{I} = \textbf{A}\delta_s^2 \textbf{I } + \delta^2\textbf{ I}$. Here we get $H(\textbf{S})$ by applying inverse Fourier Transform of CSI.


The feature extraction section in Fig. ~\ref{fig:flowChart} illustrates the steps for processing raw data into training data. All data firstly go through the data augment \cite{sobehy2019ndr}, where noise is added both on CSI and position, the dataset size is doubled and the method is more trained in this way. Then based on which features to extract, we can first convert complex to real amplitude and smooth the data by polynomial regression then downsample to the smaller dataset. Or in another way, we can do inverse Fourier transform on channel response data to time-domain, as calculated the CM, only the upper triangle is taken since the matrix is symmetric and then the amplitude is taken as the feature. These impacts of choosing different features are discussed later.

\subsection{Validation check}

\begin{figure}[h!]
  \centering
  \includegraphics[width=0.5\textwidth]{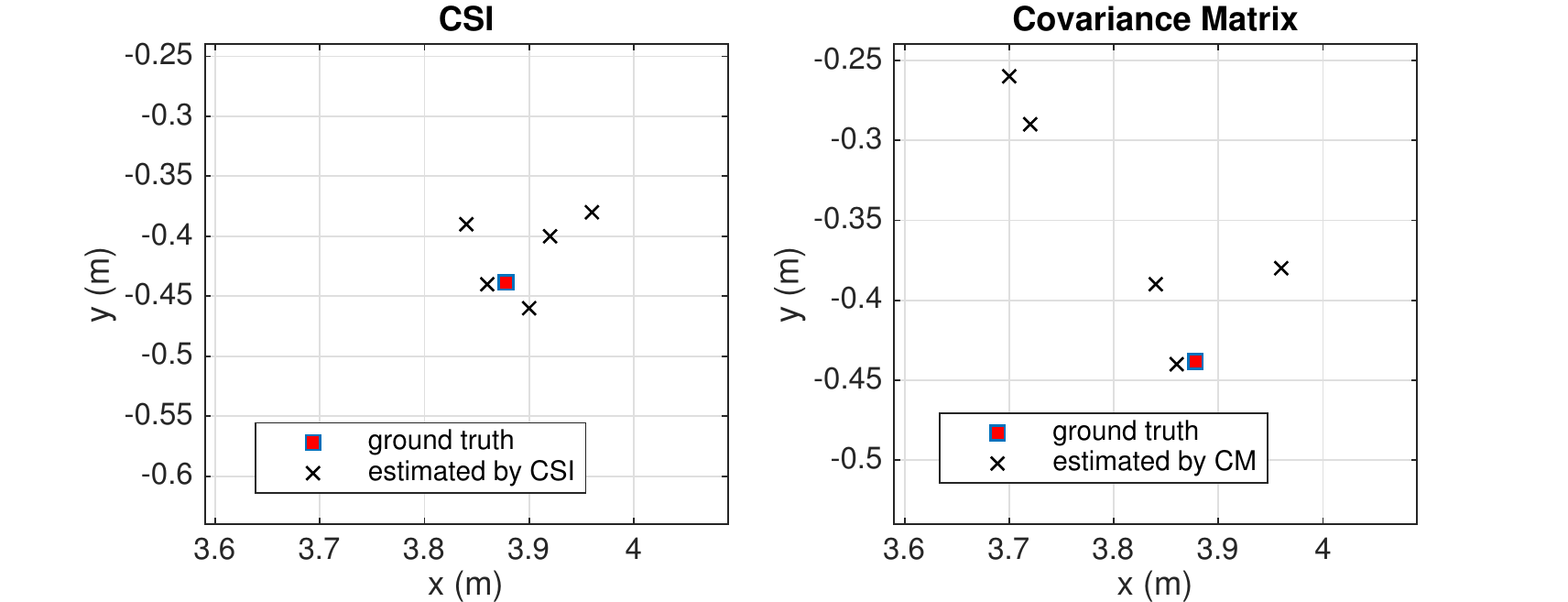}
  \caption{corresponding 5 positions of most similar features of a target feature, left use SCI feature, right use CM feature}
  \label{fig:neighbor}
\end{figure}
The common principle of the estimation method is a \textit{fingerprint matching} system, and the very basic rule of such a system is the one-to-one mapping property. Such property includes two aspects, on one hand, a similar label should have a similar feature, which is shown in previous Fig.~\ref{fig:amp_smooth} and Fig.~\ref{fig:cm}, CSI and CM both satisfy the property. It is notably in Fig.~\ref{fig:cm}, the diagonal values are not the same, which means the power level is not the same overall antennas. On the other hand, a label has to associate with a unique feature, namely \textbf{fingerprint}, so when a new feature comes in we can estimate the label by referring to these neighbors with the most similar features. Fig.~\ref{fig:neighbor} demonstrates the property, where a similar feature brings a nearby position, and CSI does a better job than CM in such a canonical method. The CSI brings can approximate bring nearby points, yet the CM gives few estimated points farther than expected.

\subsection{Small dataset}

\begin{figure}[h!]
  \centering
  \includegraphics[width=0.5\textwidth]{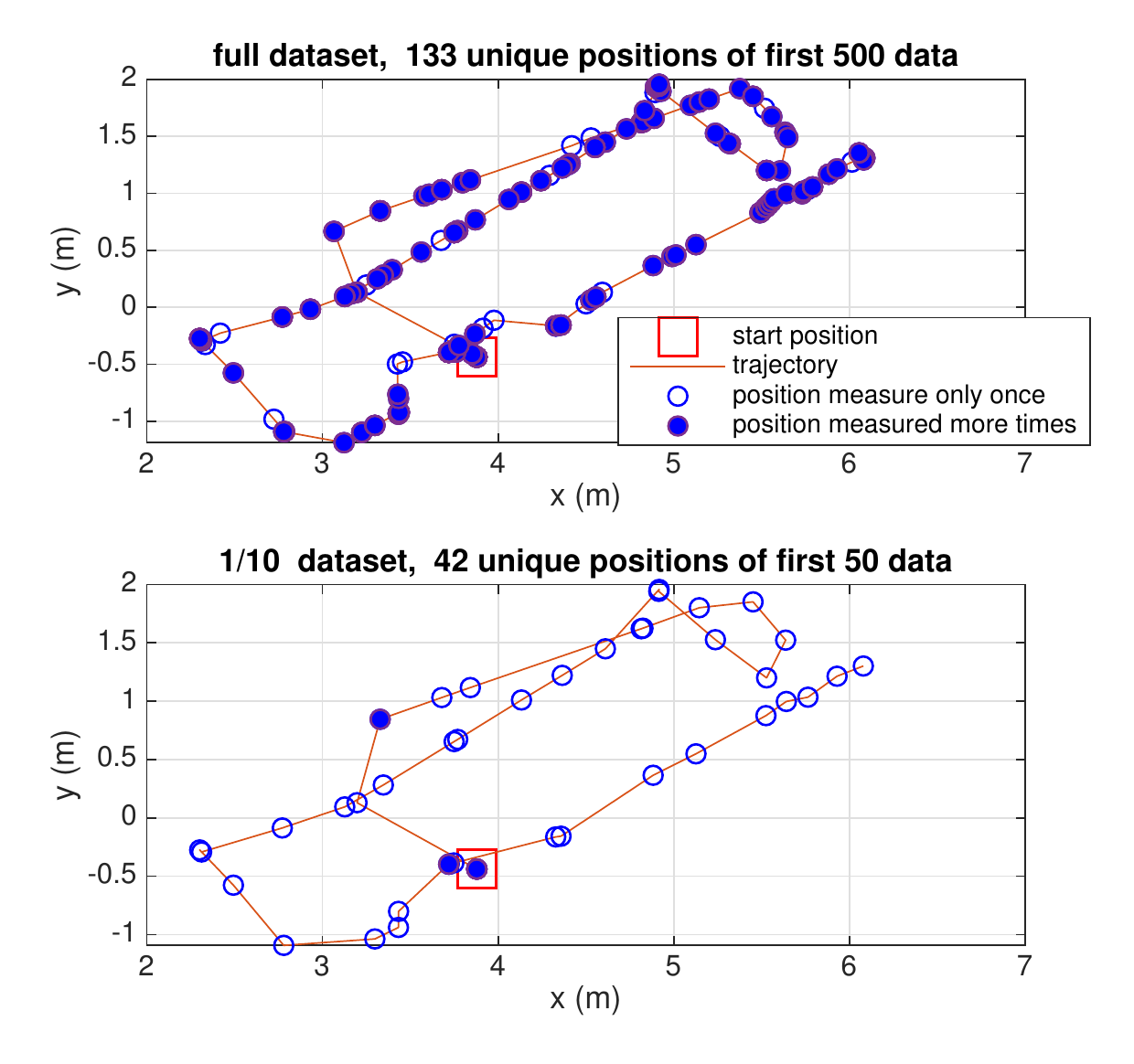}
  \caption{Full and sampled trajectory of the robot, where there are lot of position measured more than once, while less in downsample data }
  \label{fig:traj}
\end{figure}

To make a comprehensive comparison between these methods, we are also interested in small dataset learning. Although in our case of a robot moving on a desk, measured points are so close to each other and they are even measured more than twice at the same position. As shown in Fig. ~\ref{fig:traj}, we can see lots of positions are measured more than once, while after downsampling of rate 10, most of the positions are only measure once. In practice, the number of measures points are limited and they can be sparsely distributed.


\section{Positioning Estimation Method}

\begin{figure*}
\centering
  \includegraphics[height=9cm]{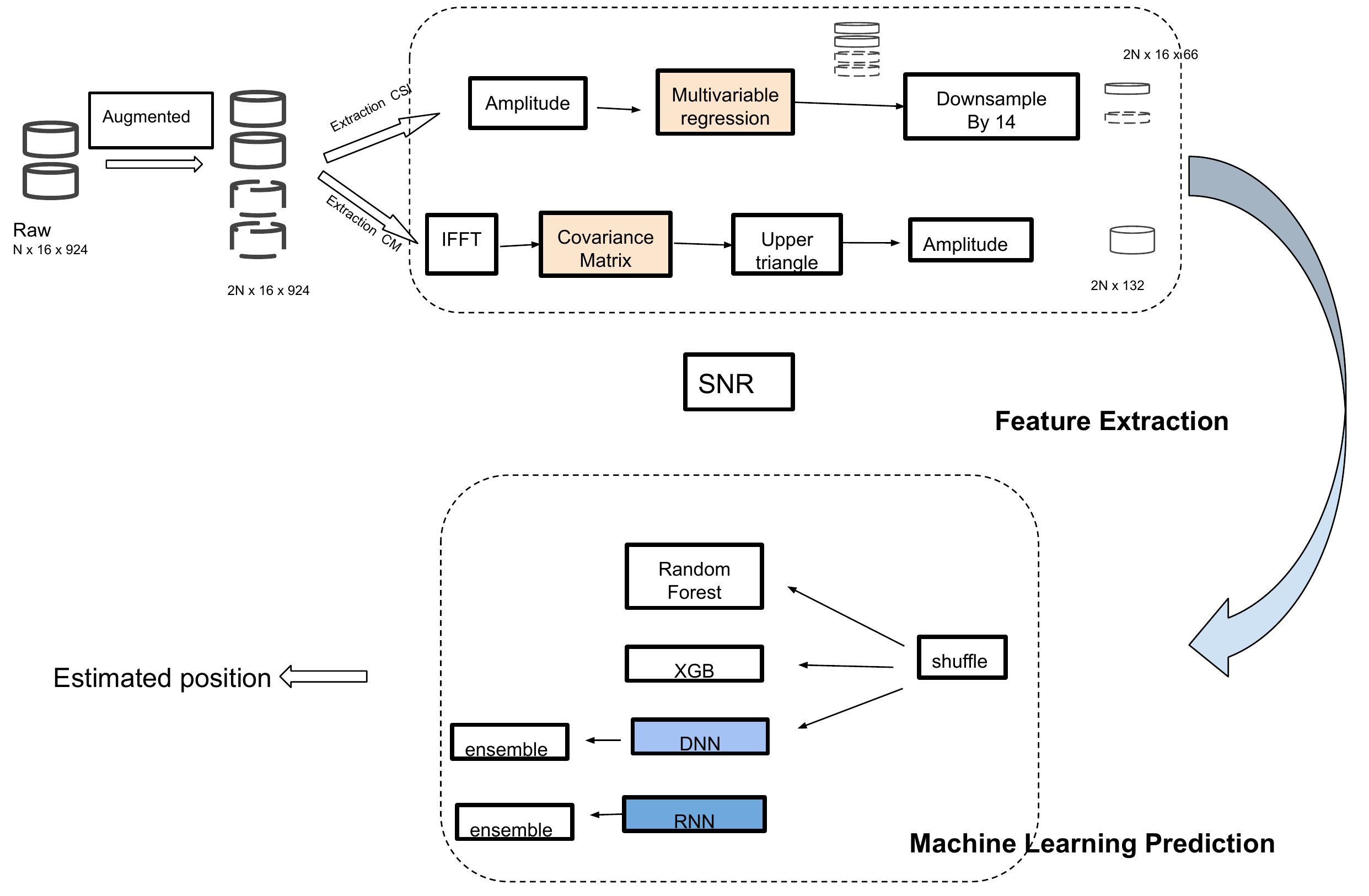}
  \caption{Flowchart of data pre-processing and training}
  \label{fig:flowChart}
\end{figure*}

At the training stage in Fig~\ref{fig:flowChart}, 4 methods are evaluated here, the random forest and XGB, the DNN and RNN. Notice that random shuffle is required for the first three, which RNN needs this trajectory information as data are arranged by the robot trajectory. Also ensemble neural network structure is applied on both DNN and RNN to better estimate the position based on differently trained NNs respectively, while decision-based methods get ensemble technics themselves.

\subsection{Random Forest (RF)}
One branch of supervised learning is the decision tree, and it estimates in a tree-like structure, where a leaf denotes a group of features and the branch represents the weight or probability, and it applies recursive binary splitting to further split grouped features into smaller combinations. Moreover, the weights are updated and some leaves are cut off by split-cost function. One of the decision tree methods is the random forest, where each tree gives a classification result and the forest chooses the class that has the highest votes. It becomes popular since it does not need for pruning trees, not sensitive to outliers in training data and can infer variable importance ~\cite{jedari2015wi}. 

\subsection{Extreme Boost Tree (XGB)}
As another decision-tree based method, XGB  ~\cite{chen2015xgboost} is the most efficient one so far, it is widely accepted in most data mining competitions of Kaggle and usually outperforms neural network when the dataset is skewed. Instead of weigh-sum over input variable in neural networks, tree learning describes the approximate function by
\begin{equation}
    \hat{y_i} = \sum_{k=1}^{K} f_k(x_i) , f_k \in \mathbf{F},
\end{equation}
where K is number of trees,and $\mathbf{F}$ for space of regression trees. The optimization objective is to minimum to training loss and complexity, 
\begin{equation}
    obj = \sum_{i=1}^{n} l(y_i,\hat{y_i}) +  \sum_{k=1}^{K} \Omega (f_k),
\end{equation}
the first item measure how well the model fit on training data and the second the item measures the complexity of trees. To generate the forest, namely tree ensemble, the algorithm provides an additive training solution, which starts from the constant estimate and adds a new function each time. Moreover, which tree to add at each step is decided by the Taylor expansion of objective. The algorithm will stop until it reaches a minimal objective goal. XGB is a self-contained derivation of the general gradient boosting algorithm, and require much less computation than neural networks. XGB only supports single value regression so several models are required to estimate all angles. XGB has several key options related to its performance. The maximum depth of each decision tree, the maximum number of training iterations until the training stop, the $\eta$ adjust the minimum training step for iteration and the evaluation metric decide the loss function.

\subsection{ Neural Networks}

Here Multiple Layer Perceptron is adapted as a network structure, the input is channel estimation and SNR, and refined input is a CSI or CM.  Several techniques applied here can help the NN train better and faster: (1), Early-stop: to prevent overfitting, which means the trained model corresponds too closely to a particular set of data, yet fail to fit another dataset reliably. Early-stop cuts a validation dataset from the training dataset, and uses it validate the error rate in each epoch, then when the neural network is overfitted, the validate error diverges from the training error and goes up, the early stop is activated and the training stop. It is notable early stop also makes the NN robust to layer and size parameters, which make the tuning easier in practice. (2), Dropout:  to ignoring units during the training phase of a certain set of neurons which is chosen at random. Dropout forces a neural network to learn more robust features that are useful in conjunction with many different random subsets of the other neurons, at the cost of larger iterations to converge.  (3), Batch Normalization: mitigate the problem of internal covariate shift, where parameter initialization and changes in the distribution of the inputs of each layer affects the learning rate of the network. (4) Ensemble NN: to construct different NNs models trained with different datasets and parameters, and estimate final value jointly by all these predictions.

\begin{figure}[h!]
  \centering
  \includegraphics[width=0.5\textwidth]{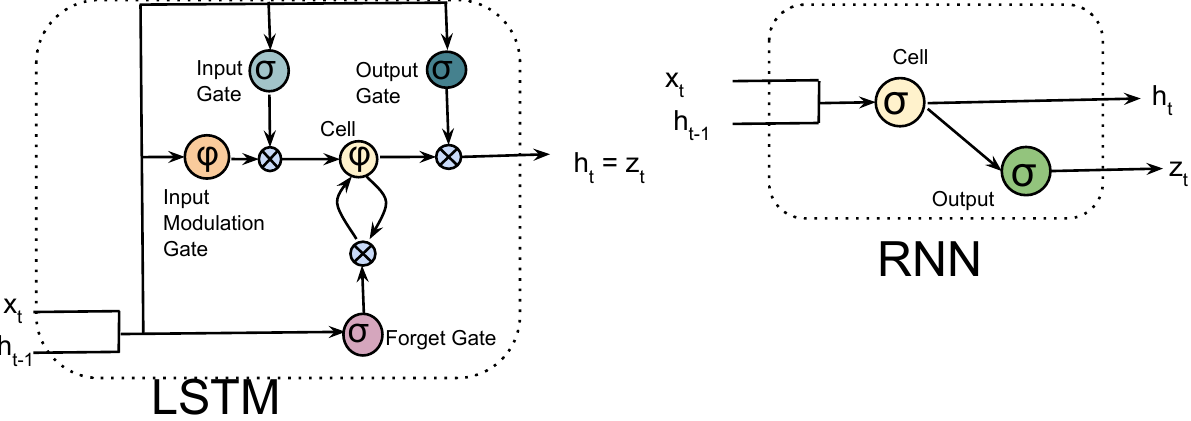}
  \caption{Illustration of a LSTM cell and  how memory flows in RNN }
  \label{fig:LSTM}
\end{figure}
The standard feed-forward deep neural network (DNN) assumes all inputs to the network and all outputs are independent of each other. Currently the input only makes use of the information over the space, while not over time. However, as we find the robot is not moving too fast, the location between each entry is quite adjacent. To this end, we propose RNN to leverage the sequential nature of the input. In this scheme, the current output not only depends on the inputs, but it also depends on the previous output much like having a memory of the previous states. As depicted in Fig. ~\ref{fig:LSTM}, the long short-term memory (LTSM) model of RNN is used, which adds input, output, and forget gate to each unit, while each gate is decided by corresponding weight trained over time.

\section{Performance Evaluation}
We use the open-sourced raw dataset, and process it both by polynomial CSI and CM, then feed the features into different methods. Table.~\ref{table:settings} lists key parameters for all methods, the DNN is in 5-layer structure with dropout layers between them, and optimizer is \textit{adam} with loss function defined as the mean square error. For RNN, the LSTM cell number is 128, with other settings the same as DNN. Both RNN and DNN apply ensemble NNs up to 10, and the final estimation is based on picking the median value. Max depth is 30 for both random forest and XGB, and XGB is tuned with minimum step 0.30 and the number of the round as 800. We only pick 10 degree for polynomial smoothing and 10 NNs for ensemble NN, instead of exhaustively searching for the best setting in ~\cite{sobehy2019ndr}. All of our dataset and codes are open-sourced at Github \footnote{https://github.com/yujianyuanhaha/Positioning} with detailed descriptions.

\begin{table}[h]
\caption{Machine Learning Settings}
\label{table:settings}
\begin{center}
\begin{tabular}{|c|c|}
\hline
Setting & Value\\
\hline
\hline
Test ratio & 0.10\\
\hline
Validation ratio & 0.10\\
\hline
Data auguemnt rate & 1.00\\
\hline
ensemble NN count & 10\\
\hline
ensemble NN selection  & median\\
\hline
Early stop patient count & 5\\
\hline
Batch Size & 1024\\
\hline
Hidden layers & [1024 512 256 128 128]\\
\hline
Drop out ratio & 0.001\\
\hline
Batch normalization L2 regularizers & 0.001\\
\hline
Optimizer & \textit{Adam}\\
\hline
Loss function & \textit{mse}\\
\hline
\hline
RNN LSTM cells & 128 \\
\hline
Random Forest depth & 30\\
\hline
XGB maximum depth & 30\\
\hline
XGB minimum step $\eta$ & 0.30\\
\hline
XGB number of rounds & 800\\
\hline
\end{tabular}
\end{center}
\end{table}

\begin{figure}[h!]
  \centering
  \includegraphics[width=0.5\textwidth]{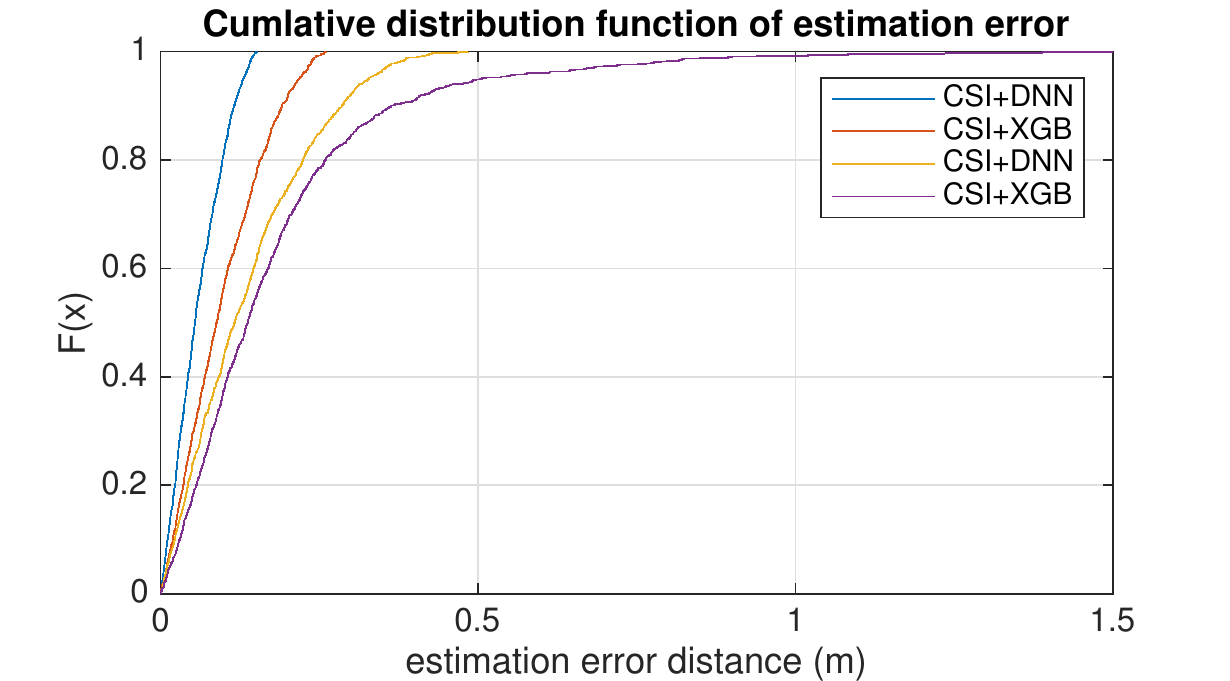}
  \caption{Cumulative distribution function of estimation error}
  \label{fig:roc}
\end{figure}

\begin{figure}[h!]
  \centering
  \includegraphics[width=0.5\textwidth]{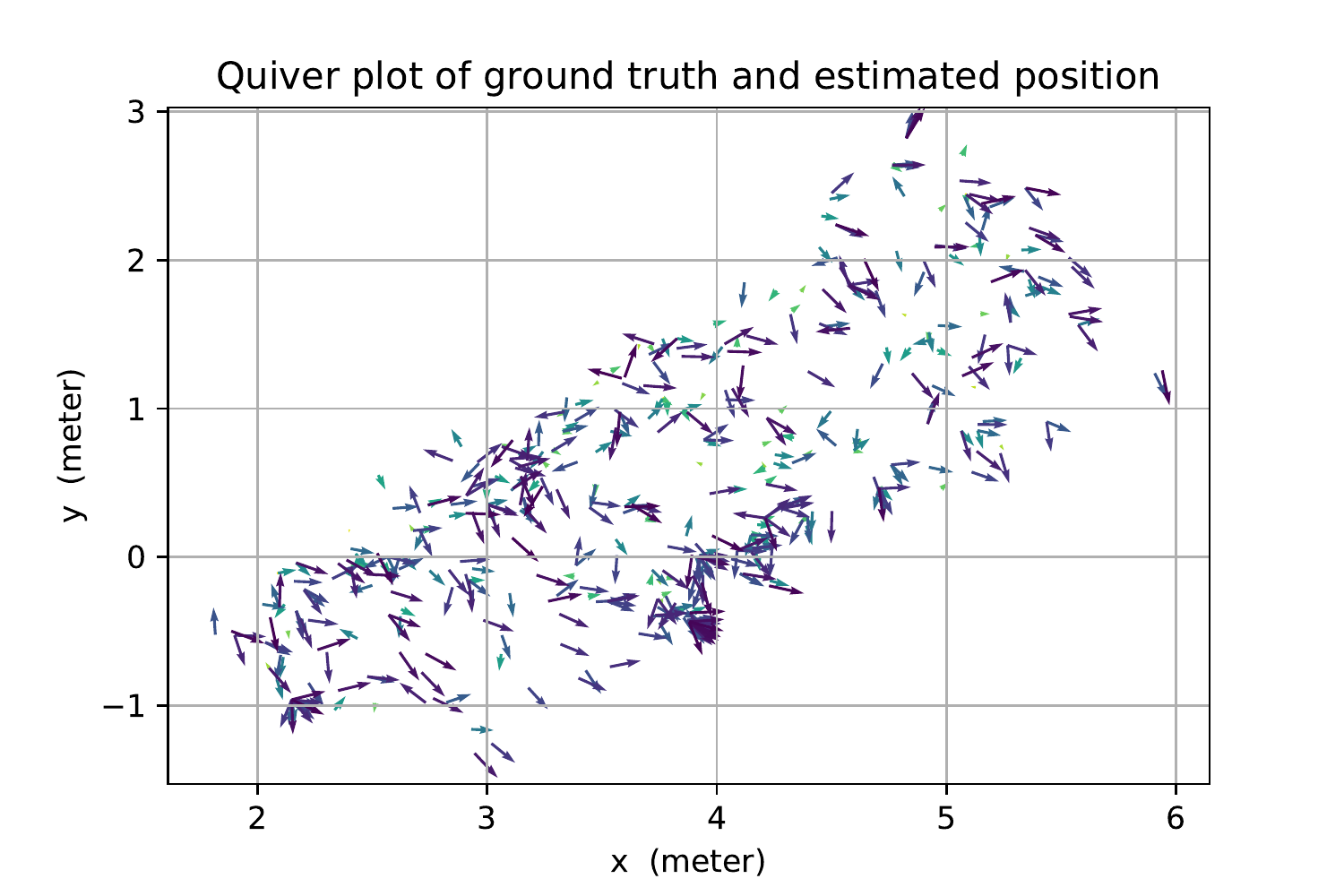}
  \caption{Quiver figure of training CSI with DNN, the start point of each  arrow is the ground truth position, while the end point denotes the estimated position, the longer the arrow, the larger the estimation error}
  \label{fig:quiver}
\end{figure}

\subsection{Full Dataset}

\begin{figure}[h!]
  \centering
  \includegraphics[width=0.5\textwidth]{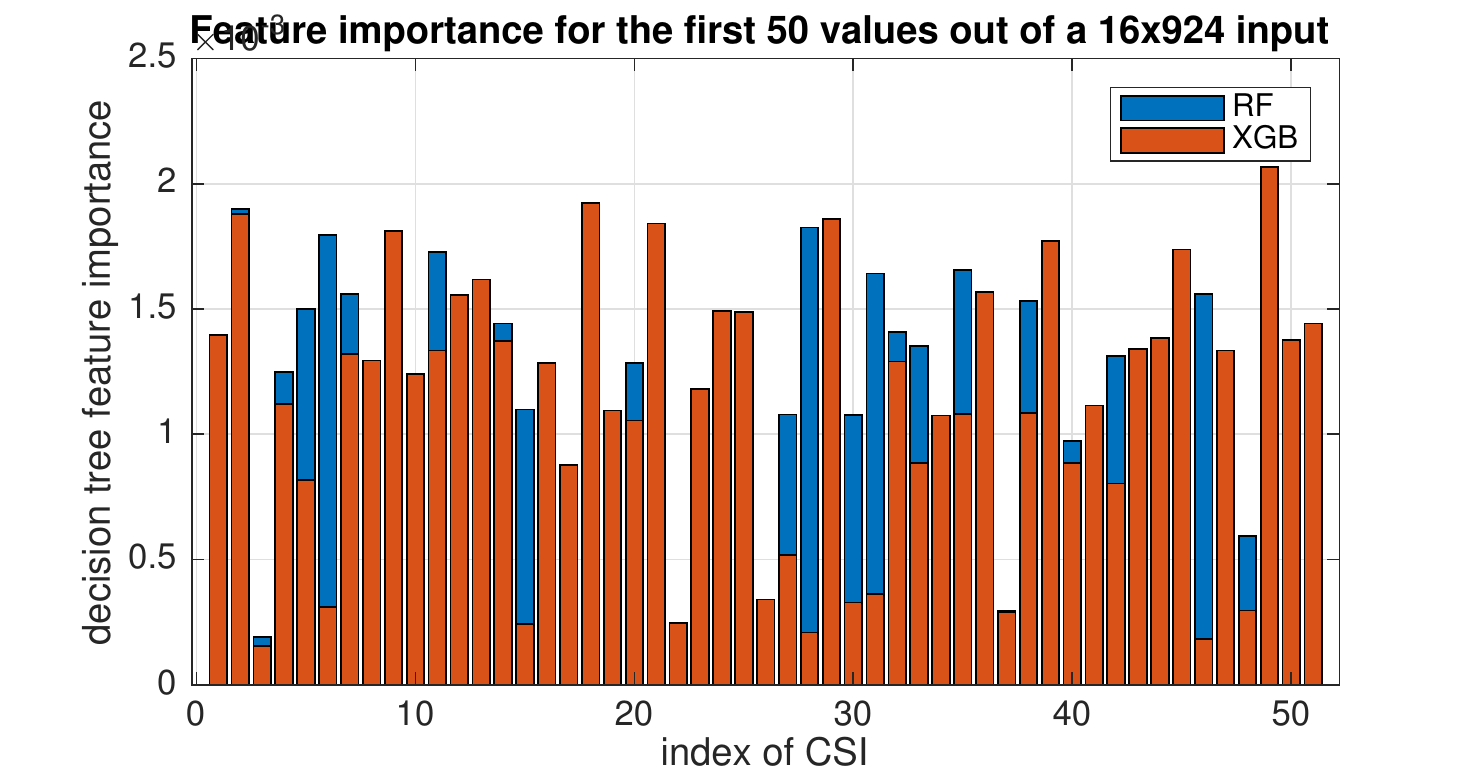}
  \caption{Feature importance of RF and XGB after training}
  \label{fig:tree}
\end{figure}

Fig.~\ref{fig:roc} depicts the performance of DNN and XGB on the feature of CSI and CM. It is shown training with CSI is significantly better than CM, while in each case, DNN is always better than RF and XGB. The exact mean absolute estimation error is shown in Tab.~\ref{table:settings}. Moreover, We found in Fig.~\ref{fig:tree}, the feature importance, an important indicator indicates how useful or valuable each feature was in the construction of the boosted decision trees within the model, varies a lot over samples of CSI. That means the decision trees learn to trust less on some elements during the training, whereas each element in a CSI measurement matters and should bring similar importance. That may help explanin why RF and XGB may miss some information and performance worse than neural networks.

\begin{figure}[h!]
  \centering
  \includegraphics[width=0.5\textwidth]{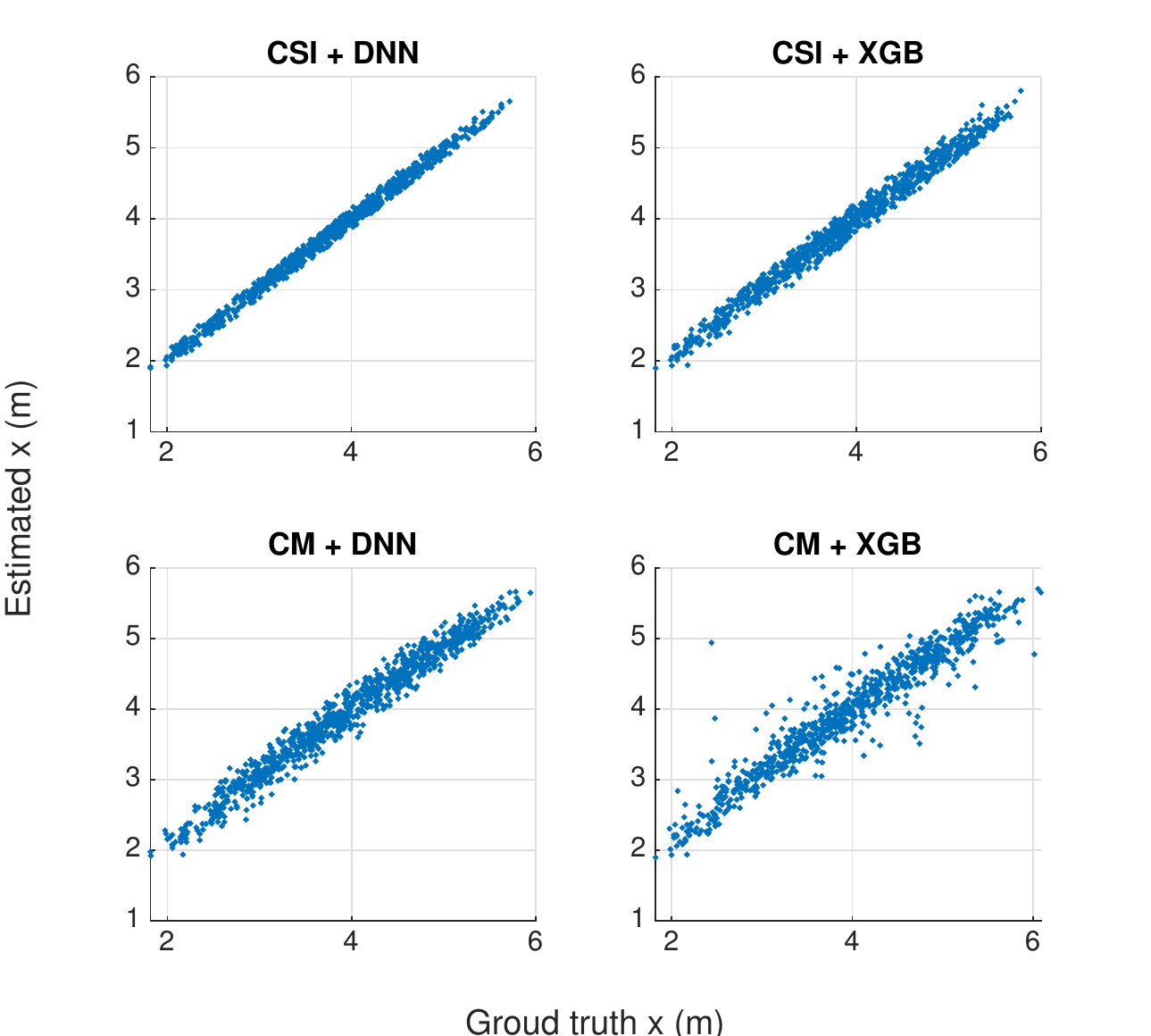}
  \caption{Scatter plot of \textit{x}-axis grouth truth versus estimation}
  \label{fig:scatter}
\end{figure}

Meanwhile, the scatter plots in Fig.~\ref{fig:scatter} also comply with the conclusion, the result of training CSI with DNN is dense on a clear straight line while CM-based  XGB brings a more sparse distribution. Fig.~\ref{fig:quiver} shows the error distribution  over different locations. The longer or darker the arrow, the higher the error. Note that for better visualization, the physical length of the arrow is proportional to estimation error not represent as the true error, which is actually much smaller. We found the error is related to the ground truth positions. The high error mainly comes from: (1). The edges of the whole trajectory, where the positioning target does not frequently measured. (2).  The faraway area from the linear array located at the upper-right part, in that channel model becomes more complicated over such long distances. It is expected if the linear arrays are placed on the ceiling of the room, the errors are less dependent on the distance between antennas and objects.

Moreover, we found several reasons why CM-based method's performance is worse than that of CSI-based. Firstly, the indoor multi-path makes the channel models complex and CM may not capture them very well. Secondly, the relatively limited view in our training and testing scenario, where a robot moves from one side of the desk to another, means the elevation and azimuth angles may not change much during the whole movement. Thirdly,  the uneven transmitting power of different antennas that makes the CM imprecise. However, CM can still provide a rough estimation based on much smaller data size with only 12.8\% of smoothed CSI, which could be a benefit in some applications.  Moreover, comparing to CSI requiring preambles to calculate the response,  CM only requires raw received data without know any preambles, which makes it more flexible in some cases.

\subsection{Small dataset learning}

Another fact we evaluate here is small dataset learning, when data is insufficient due to low-duty cycling data collection or the robot is moving fast. The small dataset comes from the full dataset downsampled a rate of 10. As shown in Table ~\ref{table:result}, with a small dataset learning, RNN performance is significantly better than DNN since the LSTM can better make use of the time consistency of the user trajectory. It is notable the test dataset is sorted by their trajectory or measured time, the RNN performance will degrade once test data are shuffled and time information is distorted.

\begin{figure}[h!]
  \centering
  \includegraphics[width=0.5\textwidth]{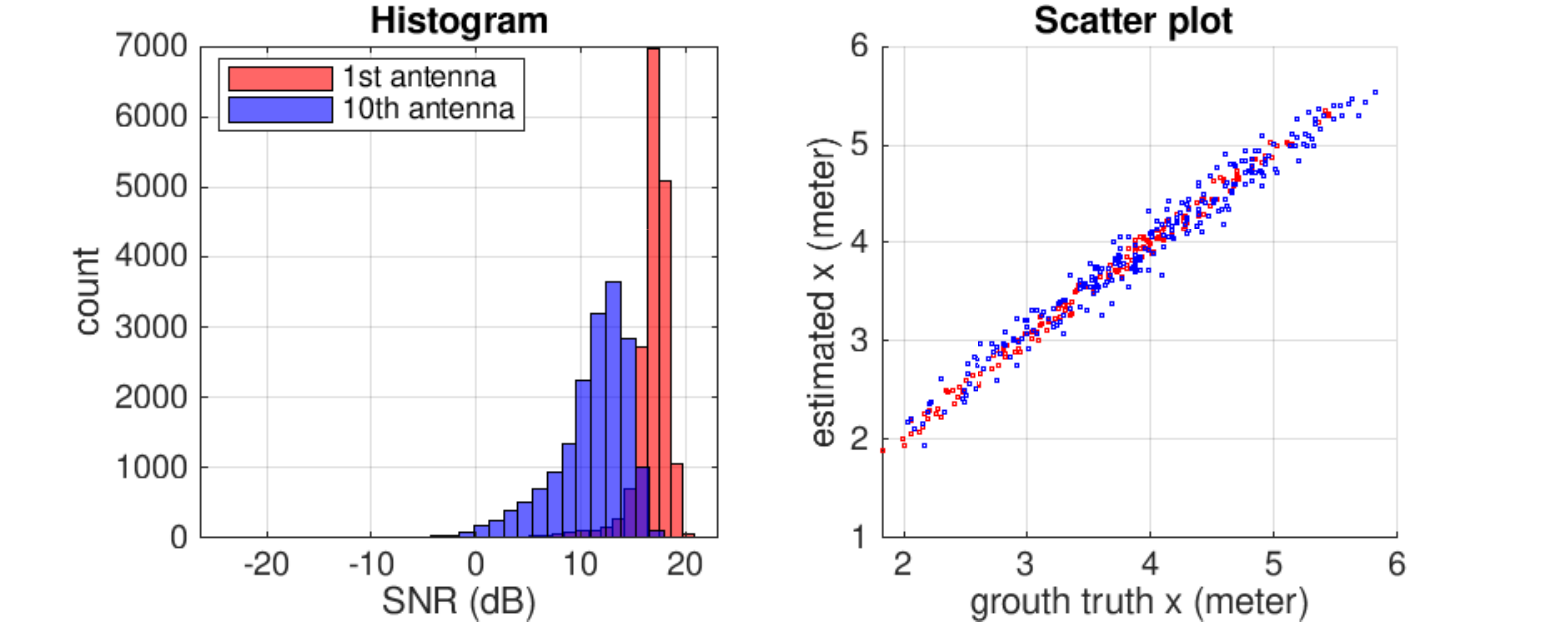}
  \caption{Left is SNR histogram of 1st and 10th antenna. Right is the NN estimation based on CSI on that attenna}
  \label{fig:SNR}
\end{figure}

What's more, it is also found that including SNR as an input can improve the estimation accuracy for all methods, in that learning method learn to trust more on high-SNR data rather than low-SNR data during the training. As shown in Fig.~\ref{fig:SNR}, SNR varies much over different antennas due to the power level, the 1st antenna has significantly higher SNR than that on the 10th antenna. Therefore, as depicted in the right plot, the NN estimation model will give a more accurate estimation on high-SNR data rather than low-SNR ones. Besides, Tab. ~\ref{table:result}  matches the conclusion.

\begin{table}[h]
\caption{Estimation error (in centimeter) of different datasets and methods}
\normalsize
\label{table:result}
\begin{center}
\begin{tabular}{|c|c|c|c|c|}
\hline
Dataset/ Method & DNN & RNN & RF & XGB\\
\hline
full CSI  & \textbf{4.55} & 9.26 & 6.67  & 6.02 \\
\hline
full CM  & 16.01 & 18.21 & 22.41  & 16.69 \\
\hline
small CSI  & 13.35 & \textbf{9.92} & 16.33  & 14.97 \\
\hline
small CSI \& SNR  & 11.95 & \textbf{8.15} & 13.50  & 12.87 \\
\hline
\end{tabular}
\end{center}
\end{table}

\section{Conclusions and Further Work}
We found that (a). The neural networks can better estimate position than decision-tree based methods. (b). Smoothed CSI can better represent as identical features than CM in indoor multiple-path cases. (c) RNN can bring higher accuracy than DNN by using user trajectory at the small dataset. (d). Leveraging SNR as input can assist the estimation.

Since more open-source CSI mining tools have been released, our further works include more subcarriers in CSI mining to increase the position accuracy,  and evaluating multiple target positioning. Besides, we will further explore CSI-mining methods in some other indoor scenarios with obstacles or interference. Meanwhile, it is also interesting to extend CSI mining for many other applications like user number count, user activity recognition, traffic identification, etc..

\addtolength{\textheight}{-12cm}  

\section*{ACKNOWLEDGMENT}

The authors thank Dr. Maximilian Arnold for providing the raw dataset and sharing the prototype code. The authors would also like to thank Virginia Tech Applied Research Corporation (ARC) for providing powerful computational machines to quickly verify our ideas.

\balance
\bibliographystyle{IEEEtran.bst}
\bibliography{IEEEabrv,references.bib}

\end{document}